\documentclass[pre,twocolumn,floatfix]{revtex4}
\usepackage{graphicx}
\usepackage{epsf}
\usepackage{times}

\begin{document}

\newcommand{\muAl}{\mu_{\rm Al}}
\newcommand{\Al}{_{\rm Al}}
\newcommand{\AlAl}{_{\rm Al-Al}}
\newcommand{\AlCo}{_{\rm Al-Co}}
\newcommand{\CoCo}{_{\rm Co-Co}}
\newcommand{\aR}{{a_R}}
\newcommand{\rcut}{{r_{\rm cut}}}
\def\OMIT#1{}
\def\WISHLIST#1{}
\def\MEMO#1{}
\def\NOTE#1{}
\def\CHECK{{\bf [CHECK] }}

\title{Matching rules from Al-Co potentials in an almost realistic model}

\author{Sejoon Lim}    
\altaffiliation[Current address: 
311-602 Hyundai Apt.;
Bora-dong, Giheung, Yongin;
Gyeonggi-do, South Korea 446582]
{}
\author{Marek Mihalkovi\v{c}}
\altaffiliation{Also Institute of Physics, 
Slovak Academy of Sciences, Dubravska cesta 9,
84511 Bratislava, Slovakia (permanent address).}
\author {Christopher L.  Henley}    
\affiliation{Dept. Physics, Cornell University,
Ithaca, New York, 14853-2501}


\begin{abstract}
We consider a model decagonal quasicrystal of composition Al$_{80.1}$Co$_{19.9}$
-- closely related to actual structures, and using realistic pair 
potentials -- on a quasilattice of candidate sites.
Its ground state, according to simulations, is a Hexagon-Boat-Star tiling
satisfying Penrose's matching rules. 
In this note, we rationalize these results in terms of the potentials;
the Al-Co second-neighbor potential well is crucial.
\end{abstract}

\maketitle


From the discovery of quasicrystals, it has been difficult to
decide whether they are stabilized by entropy or by energy.
(Here ``energy'' means an ideally quasiperiodic ground state.)
\WISHLIST{Diffuse scattering near Bragg peaks suggests the entropic
answer, but only in icosahedral quasicrystals.}
In the decagonal case,
the fastest route to a resolution may be ab-initio-based modeling.
Recently, we observed a ground state~\cite{QC07} that perfectly 
implements Penrose's matching rules~\cite{penrose} in 
a toy model of a binary Al-Co quasicrystal,
closely related to realistic d(AlNiCo) models~\cite{mihet02,Gu-pucker,Gu-PML}.
In this contribution, we sketch the matching rules'
origin; in particular, how fine-tuned must the potentials be to obtain this result?


Our simulations use a recipe introduced to
study real Al-Co-Ni phases~\cite{mihet02,Gu-PML,Gu-pucker} 
\OMIT{at the Ni-rich and Co-rich ends.}
\OMIT{These turned out to approximate
the structural energy quite well, as calibrated
against many ab-initio computations.}
The only input data are 
the number density, the composition ratio, and
the (quasi)lattice constants
(tile edge is $\aR \equiv 2.455$ \AA{} and layer
spacing $c/2\equiv 2.04$\AA).
In an initial ``unconstrained'' Monte Carlo simulation
the atoms hop as a lattice gas~
\OMIT{\cite{Cockayne1998}}
on discrete, properly placed candidate sites (see ~\cite{mihet02,Gu-PML}), 
which decorate random rhombus tilings (that are rearranged as
another kind of Monte Carlo move).
The {\it lowest} energy configuration
visited during the run is saved (which  is
far better than a typical state, due to energy fluctuations 
in these relatively small systems).
The second stage is a ``constrained''
simulation flipping HBS tiles with a fixed atomic decoration
(which allows larger systems)
inferred from the unconstrained results.
\OMIT{The atoms deterministically decorate the HBS tiles, 
so the only degrees of freedom are tile flips; in 
effect we have a random-tiling simulation, with a 
tile-tile Hamiltonian implicitly given by the potentials.}


\begin{figure}
\includegraphics[width=3.4in]{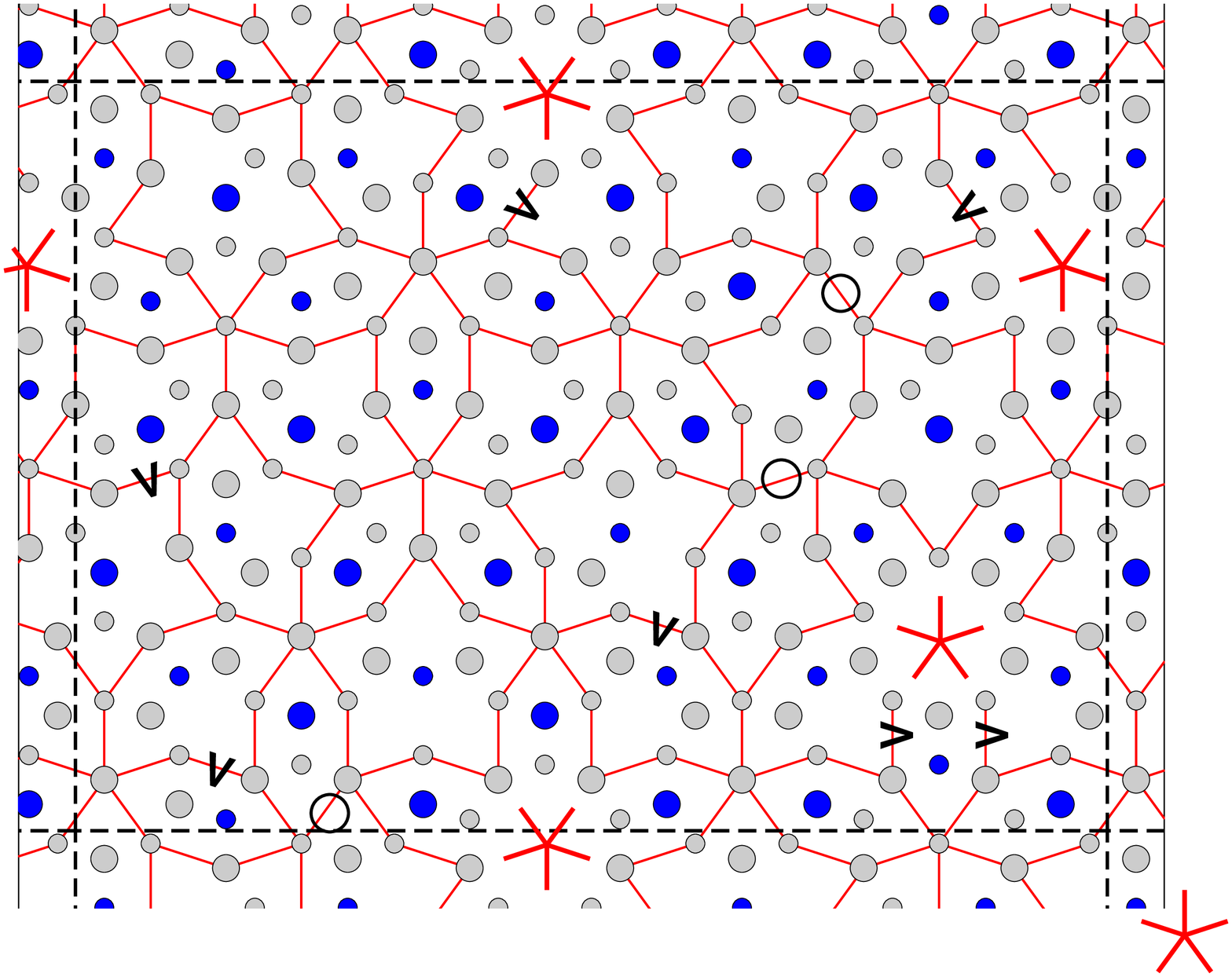}
\caption{Best unconstrained configuration from original 
discovery runs, 
placing atom content Al$_{169}$Co$_{42}$
in a smallish cell 31.9 $\times$ 23.3~\AA, 
\OMIT{$\tau^5 (1.172) \aR \times \tau^3 \sqrt{5} \aR}
with periodic boundary conditions.
We cooled gradually to $T\approx 1100$K 
with $10^5$ trial atom swaps per swappable pair, 
and 2000 tile flips per flippable pair during the run.
Co and Al atoms are shown by
black and gray circles; large/small circles
indicate the top/bottom layers.
\OMIT{For this composition, the atoms 
organize into a perfect Penrose tiling, with 
several matching-rule violations.}
The seven V-rule violations (i.e. fat rhombus/thin rhombus) are marked ``V'', 
the three fat/fat violations  are marked with circles,
and the three vacated Al sites are marked by asterisks.
Four violations are necessitated by periodic boundary condition.
The pair of V's in the lower right corner mark 
an energetically favored defect (see text).}
\label{fig:bestAlCo}
\end{figure}

The result (Fig.~\ref{fig:bestAlCo})
is a Hexagon-Boat-Star (HBS) tiling of edge 2.45~\AA~
with Al atoms at exterior vertices 
and Co atoms on the interior rhombus vertex, as well
as (respectively) one, two, or two internal Al atoms.
Importantly, an Al vertex site 
goes vacant if it would have no Co nearest neighbors.
\OMIT{this happens on vertex type $\la 22222 \ra$ of the Penrose tiling.
i.e. where surrounded by a Star tile of the 4.0-HBS tiling.}
For the present work, we introduced a ``half-constrained'' simulation
in which atoms hop as a lattice gas on HBS tiles, limited to a site list
of the ideal vertex sites, plus sites $1/\tau^2$ out along the
midline of each Fat rhombus.
This revealed that rearrangements are
induced next to the empty vertices, such that the ground state is not literally
the Penrose tiling (e.g. the defect marked by a pair of V's in Fig.~\ref{fig:bestAlCo}.)
However, we believe quasiperiodicity is maintained (the shape of hyperatoms 
in 5D space will change slightly).

\OMIT{The deterministic decoration rules do not
allow us to remove Al from the special sites lacking Co neighbors.}

\WISHLIST{
The atom sites in the simulation correspond to the
perfect occupation domains in ``perpendicular'' space 
(actually, we expect partial disorder in the orientation 
of the two internal Al in some Star tiles).}

\WISHLIST{The matching-rule structure is stable for the broad interval 
$-2.05 < \muAl < -1.55$ eV.
(Equivalently, this is the energy per HS $\to$ BB tile flip.)}

\vskip0pt plus2mm
\section{Ordering and diagnostics}
\label{sec:order}

The interactions (we used ``GPT'' potentials~\cite{mihet02,Moriarty97})
have the following properties:
(i) $V_{\rm AlAl}(R)$  has a strong hardcore repulsion 
at close neighbor distances ($R<2.8$\AA). 
(ii) $V_{\rm AlCo}(R)$ is extremely attractive at
$R\approx2.5$\AA;
(iii) $V_{\rm CoCo}(R)$ 
has a rather strong attraction at the second-neighbor distance $R\approx4.5$\AA,
due to the prominent Friedel oscillations;
(iv) similarly $V_{\rm AlCo}(R)$ is attractive, too, around $R\approx 4.5$\AA;
this well is mainly responsible for the matching rules~\cite{QC07}.

How does this order follow from these potentials?
The question falls into two parts.
(i) Why the HBS tiling?  
(ii) Granted the atoms do form an HBS tiling, why do
the interactions force matching rules?  
This is the focus of Sec.~\ref{sec:mrules}, which
uses inflated (edge 4~\AA) HBS tiles to examine the
energies  with more detail than Ref.~\cite{QC07}.

In Ref.~\cite{QC07}
we identified four aspects of ordering, 
the first two of which develop at relatively high temperatures
($T\ge T_1 \equiv 1000$ K).
Aspect (1) is formation of the 2.45-\AA~  HBS tiling,
with Co in the center and Al decorating the corners of every tile,
as is quite universal in Al-transition metal decagonals.
HBS tiles trivially satisfy the Penrose {\it double} arrows.
\OMIT{they point toward the center vertex decorated with Co.}
The HBS formation is attributed to the nearest-neighbor
Al-Al and Al-Co interactions~\cite{QC07}.
Medium-temperature configurations 
from an ``unconstrained'' simulation may have 
many defects (of matching rules, or placement of Al atoms
internal to the HBS  tiles), but the HBS framework itself
is inviolate.

Aspect (2) is the Co-Co network due to second-neighbor Co-Co attraction:
a supertiling  of inflated HBS tiles 
(edge $\tau a_R \approx 4$ \AA) 
and small generalizations (e.g. pillow tile, below).  But at medium $T$, the
Co-Co tiling is not yet a Penrose inflation of the 2.45\AA-level HBS tiling.

\WISHLIST{
The arrangements of Al within the three HBS tiles
seem to be the three best ways of maximizing the number of the (very strong)
Al-Co nearest-neighbor bonds while minimizing Al-Al repulsions
[but note some $R\AlAl=2.45$~\AA~ pairs].
}


\begin{table}
\caption{
Important bond distances and potential values,
bonds of each kind $N(R)$ per cell, and 
change $\Delta E^R_{a,b}$ in the contribution
to the total energy change $E(T_b)-E(T_a)$, per cell.
(Interlayer distances are written $R_2$.)
The same cell and atom composition were used as 
in Fig.~\ref{fig:bestAlCo}; temperatures were
$T_\infty\approx 5000$,K  $T_1\approx 2000$K (using a
randomly chosen typical configuration after
equilibration, from the ``half-constrained'' simlation); $T_0$ refers to the near
ground state, the best found during a
run at low temperature.}
\begin{tabular}{@{}*{6}{l}}
  $R$ & pair & $V_{AB}(R)$ & $N(R)$  &   $-\Delta E_{0,1}$ &  $-\Delta E_{0,\infty}$ \cr
(\AA) &  $(AB)$ & (eV)     &    & (eV/cell) &  (eV/cell)  \cr
  2.455  & Al-Al &  0.4124    &  119  &  +0.8247     &     +5.7729     \cr
         & Al-Co & $-0.2919$  &  152  &  -0.5839     &     +1.4597     \cr
  2.542  & Al-Al &  0.2758    &  248  &  -1.1033     &     -3.8615     \cr 
         & Al-Co & $-0.2583$  &  240  &  +2.0661     &     +1.0331     \cr
         & Co-Co &  0.1001    &    0  &   0.0000     &     +0.2002     \cr
  2.886  & Al-Al &  0.0791    &  170  &  -0.7119     &     -0.7119     \cr
         & Al-Co & $-0.0427$  &   52  &  +0.0427     &     +0.3416     \cr
         & Co-Co &  0.0950    &    0  &  +0.0000     &     +0.3801     \cr
  3.016  & Al-Al &  0.0669    &    0  &  +0.0669     &     +0.0000     \cr
  3.192  & Al-Al &  0.0637    &  274  &  +0.6373     &     +1.5296     \cr
         & Al-Co &  0.0913    &    0  &  +0.9127     &     +0.0000     \cr
  3.789  & Al-Al &  0.0264    &  126  &  -0.0528     &     -0.5281     \cr
         & Al-Co &  0.0554    &    2  &  +0.5538     &     +1.4953     \cr
  3.846  & Al-Al &  0.0212    &  126  &  +0.6375     &     +0.3825     \cr
         & Al-Co &  0.0424    &    0  &  +0.0848     &     +1.2722     \cr
  4.303  & Al-Al & $-0.0055$  &   144 &   -0.0439    &     -0.0439     \cr
  4.465  & Al-Al & $-0.0071$  &   456 &   +0.3811    &     +0.4940     \cr
         & Al-Co & $-0.0349$  &   528 &   +1.3261    &     +1.3959     \cr
         & Co-Co & $-0.0909$  &   116 &   -0.5453    &     +1.2724     \cr
  4.669  & Al-Al & $-0.0049$  &   283 &   +0.0147    &     -0.0978     \cr
         & Al-Co & $-0.0260$  &   149 &   +0.0521    &     +0.8337     \cr
         & Co-Co & $-0.0796$  &    45 &   +0.3183    &     -0.1591     \cr
    4.762  & Al-Al & $-0.0028$ &   247 &   -0.1039    &     -0.2254     \cr
           & Al-Co & $-0.0189$  &   304 &   -0.0756    &     +0.1702     \cr
  4.997  & Al-Al & $ 0.0015$  &  340  &  -0.0278    &      -0.0278     \cr
         & Al-Co & $ 0.0012$  &  104  &  -0.0023    &      -0.0187     \cr
         & Co-Co & $-0.0193$   &     0 &   0.0000    &      -0.1543     \cr
\end{tabular}
\label{table:alco-pot}
\end{table}

The other two aspects of order appear at $T<T_1$
and implement matching rules.
Aspect (3) is the ``V-rule'': a convex ($2\pi/5)$ corner
of the Thin rhombus in the (2.45\AA) H or B tile can only adjoin
a concave ($3\pi/5)$ corner of the B or S tile, and vice versa.
This ties together the HBS tiling and the Co supertiling,
requiring that every Co-Co edge have a Fat Hexagon around it.
\OMIT{Many examples are seen of tilings from $T \approx T_1$ with perfect
HBS and Co tilings, but having copious violations of the V-rule.}
The V-rule is simply the Penrose (single-arrow) matching
rule between Fat and Thin rhombi, as it is manifested on
our (small) HBS tiles.
Aspect (4) of the ordering -- the final stage -- is to 
satisfy Penrose single arrows between two Fat rhombi, on
the edges not involved in the V-rule (two on every
Hexagon and Boat); its explanation is deferred to Sec.~\ref{sec:mrules}.

Of course, the V-rule demands an equal number of the respective kinds of
corner, thus $n_V= 2n_H+n_B=2n_B+5n_S$.
\OMIT{Here $n_X$ is the density of $X$ tiles (per rhombus).}
Combined with the formula for the numbers of rhombi,
$n_H+3n_B+5n_S=n_{\rm fat}$ and $2n_H+n_B=n_{\rm thin}$, 
this completely specifies the number of H, B, and S tiles.
In the limit of zero phason strain, 
\OMIT{$n_{\rm fat}=\tau^{-1}$ and $n_{rm thin}=\tau^{-2}$}
we get densities $(n_H,n_B,n_S)=
(\tau^{-4},\tau^{-5},\tau^{-5}/\sqrt 5)$ (per rhombus) exactly as in
the Penrose tiling.  
(Also, if matching rules are obeyed,
the vacated Al sites are centers of inflated Star tiles
and their density is $n_{\rm vac}= \tau^{-7}/\sqrt 5$, which gives 
the Al$_{80.1}$Co$_{19.9}$ stoichiometry, and the point density
of 0.0697/\AA$^3$.)

\OMIT{Namely, there are $\frac{1}{2}(9+1/\sqrt{5})\tau^{-3}$ Al 
and $\tau^{-1}/\sqrt{5}$ Co atoms per Penrose tiling rhombus of edge $a_R$, 
which gives the quoted stoichiometry;
and a normal number density 0.0697/\AA$^3$.}

\OMIT{In general, 
$n_{\rm vac}$ depends on how the HBS tiles get put together. }

There were two key diagnostics  for our initial identification
of the key interactions for matching rules.
First, we compared the contributions of each pair distance 
$R$ to the total energy as a function of temperature
(Table~\ref{table:alco-pot}).
As evidence for the story just given,
big changes are seen upon cooling
$T_\infty \to T_1$, but none from $T_1$ downwards, in
(i) nearest neighbor [2.45--2.86\AA] Al-Al and Al-Co,
forming HBS tiles, and (ii) in Co-Co 4.47--4.67\AA, forming the
Co-Co network.  The big changes at lower $T$ were in Al-Co at
3.79--4.67\AA, in accord with the analysis of
Sec.~\ref{sec:mrules} (below).

The second diagnostic was to vary the potential cutoff
radius $\rcut$ (which was usually 7 \AA).
At $\rcut=5.1$\AA~ the Penrose rules are still satisfied whereas
at $\rcut\approx 3.5$ \AA~ one gets an HBS tiling but no matching
rules, confirming independently that HBS formation is due to 
nearest-neighbor bonds but matching rules come from second-neighbor bonds.
Furthermore, if $\rcut=5.1$\AA~ and internal Al's are removed
from all tiles,~\footnote{
With the internal Al removed but $\rcut=7$\AA,
the correct Fat-Fat rule is {\it still} upheld,
presumably by  Al(a)-Co interactions at $R\approx 6.5$\AA.}
the tiling implements the V-rule but not the Fat-Fat rule, 
confirming the role assigned in Sec.~\ref{sec:mrules}
to the ``$b$'' atoms in Fig.~\ref{fig:HBS-inf} and
similar Al.

\vskip0pt plus2mm
\section{Matching rules via 4.0 \AA~ tiling}
\label{sec:mrules}

Aspect (4) of the ordering -- the final stage -- is to 
satisfy Penrose arrows on the remaining Fat rhombus edges (two on every
Hexagon and Boat).
\OMIT{If we just knew the energy is 
higher when two Fat rhombi are parallel, that would give
the matching rule and we would be done.}
Let's focus now on the Fat rhombus/Fat rhombus matching rule, 
the crucial aspect (4) of the ordering process.
To check the energy cost of a Fat/Fat violation, which depends on the
internal Al in those Fat rhombi, 
We reformulate the matching-rule problem in terms of the 
4.0\AA~ {\it supertiles}, in which arrows are defined on 
every edge by the Al atom position. 
\OMIT{The 4.0\AA-HBS formulation is convenient for 
considering longer-ranged interactions
($5.1$\AA~ $< R < 7$\AA,
or between tiles that don't share an edge)
and for implementing certain constraints.}
A given supertiling can be
broken into HBS tiles in several ways, corresponding
to different arrowings of its edges.
Note that usually (as in our 2.45\AA-HBS tiling),
the degrees of freedom are in the tiles, and possible mismatches occur
along tile-tile edges;  whereas in the 4.0\AA-HBS model,
each edge is an independent degree of freedom, and the
possible mismatches occur between two edges in the same tile.
(The same was true for the near-matching-rule of $d$(AlCuCo)
\cite{widom-match}.)

\MEMO{Could I save space by defining a notation ``HBS+'' to distinguish 
the super tiles?}

\begin{figure}
\includegraphics[width=3.4in]{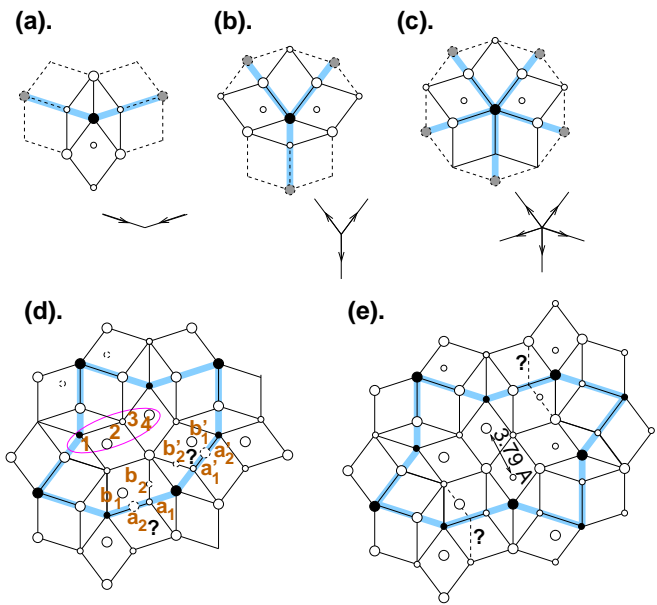}
\caption{Inflated edges (broad shaded lines),  as forced by the ``V-rule''.
Solid lines show (a) H, (b) B, and (c) S tile; dashed
lines are fragments forced by the V-rule. Insets below 
show the arrow decorations at the inflated scale
(d).  The arrowings of edges marked ``?'' are undetermined by the V-rule.
Interactions between the labeled atom sites determine
the Penrose arrowing as optimal.
(e).  Besides H, B, and S, only this pillow tile is
consistent with the ``V-rule'', but it forces the
unfavorable Al-Al distance of 3.79\AA.}
\label{fig:HBS-inf}
\end{figure}

The Co-Co network 
has 4.46\AA~ bonds, along supertile edges, and 4.67\AA~ bonds. 
All of the latter relate endpoints of a $2\pi/5$ supertile
corner (e.g. in Fig.~\ref{fig:HBS-inf}(c)).  Assuming the V-rule, 
[see Fig.~\ref{fig:HBS-inf}(a,b,c)], that makes five bonds per 
(small) B and five per small S tile.
Hence, if the H-B-S content is fixed (as implied by the V-rule), the number of
4.67\AA~ Co-Co bonds is fixed and they do not contribute to the matching rule.
(That same number, in the super tiling,  is 2 per H, 3 per B, and 5 per S tile,
so the super HBS content is also constrained.)

What supertiles are possible?
A corollary of the V-rule
\OMIT{Consider the corner angles of $2\pi/5$, $2(2\pi/5)$, or $3(2\pi)/5$}
is that, at any $(2\pi/5)$ [resp.  $3(2\pi/5)$] corner,
both arrows point out from [resp. into] the corner. 
The other constraint is that no interior space be left after
a supertile's border is decorated by HBS tiles
(filling an interior always violates the V-rule).
The conclusion is we can have the super-H, B, and S tiles, 
or the ``pillow'' tile shown in Fig.~\ref{fig:HBS-inf}(e)
(or extensions of it by adding more segments of alternating 
$2(2\pi/5)$ and $3(2\pi/5)$ corners.)

So let's consider the super-Boat tile (Fig.~\ref{fig:HBS-inf}) (d)).
The V-rule leaves undetermined whether the two edges marked  ``?''
both point to the same corner (as shown) or are flipped 
with one in and one out (this version violates the Penrose rule).
Depending on the choice for the first arrow, we either have Al at 
$a_1$ and $b_1$, or else at $a_2$ and $b_2$; similarly the 
primed sites depend on the second arrow.  The energy difference
between  the options for first arrow depends on 16 possible interactions 
between sites $a_1$, $b_1$, $a_2$, $b_2$ and the fixed atoms marked 1,2,3,4.
(Several other fixed atoms interact, but they are symmetrically placed
relative to the flip, so they can't affect the energy difference.)
In addition, there are four interactions between $(a_i,b_i)$ and $(a_i',b_i')$,
however we must exclude those that connect two sites on the same tile.
(That term was counted already in the energy per tile, and as noted earlier
the total counts of H, B, and S tiles were fixed by the V-rule.)

Here are the biggest contributions (energies are in Table ~\ref{table:alco-pot}):
either option makes one Al-Co(4.67) bond; the ``right'' way also gets two Al-Co(4.47)
-- quite favorable -- and one  Al-Al(2.89) -- very unfavorable; the ``wrong'' way
gets four Al-Al(3.85) and two Al-Al(3.79) which are unfavorable.
The ``right'' way also makes eight Al-Al(4.47) -- favorable \OMIT{but weaker} --
versus just two for the ``wrong'' way.  The net difference is 0.1825 eV
in favor of the ``right'' way.
Preliminary numerical tests, correlating total energy with the number
of matching rule violations, indicate a typical matching-rule
defect (of either kind) actually costs the order of 0.1 eV.

Next, consider the super Hexagon.
Its side edge arrows are free to point either way, but a calculation
like that for the Boat says these arrows are favored to be parallel, which
is the proper Penrose arrowing of a Boat.
Finally, the V-rule permits just one possible arrowing
on the super-Pillow tile (Fig.~\ref{fig:HBS-inf}), 
which forces a Fat-Fat rhombus violation
in the middle associated with unfavorable Al-Al distances. Hence we expect
this tile loses out to the super-Star (which carries the same count of 
2.45\AA~ HBS tiles).

\WISHLIST{Since the number of 4.67~\AA~ bonds is increased by one for every 
4.0\AA-Star tile,  why doesn't this drive the large HBS tiling to
form more Stars (and thus to a non-Penrose ratio of H:B:S)? 
The answer is, the V-rule violations cost more.
Each V-rule violation switches the distances in 
Al-Co(4.46)+Al-Al(3.79), which is a net cost  79.9 meV.
The net gain of Co-Co(4.67) is 79.4 meV.  Most important,
each flip which changes the 4\AA~ Co-Co tiling is, on
the 2.45\AA~ scale, an end-to-end flip of the 2.45\AA~ H;
this makes {\it two} V-rule violations while gaining only
{\it one} Co-Co (4.67) bond.}

\vskip0pt plus2mm
\section{Discussion}

The matching rules are mostly implemented
by Al-Co interaction at $\sim 3.8$~\AA~ and $4.46_2$~\AA;
however, they are a resultant of many terms, and the
cost of a rule violation seems to be context-dependent
(especially with our standard potential cutoff at $7$\AA,
which includes the third well.)  Fortunately, as seen 
in the example of Sec.~\ref{sec:mrules}, this isn't
a ``frustrated'' problem: most contributions have the
same signs.  We checked the cost $E_{\rm match}$ of a rule violation: 
a fit of total energy versus number of violations 
(counted by hand) in 10 low-energy configurations 
(from $T\approx 10^3$K) gave $E_{\rm match} \approx 0.1$eV.
If we took a (near) ground state and moved the violation
along a ``worm'' by local HBS flips, however, the cost
was $\sim 0.5$eV, suggesting the minimum four violations 
forced by periodic boundary conditions had found
sites where their cost was anomalously low.

\OMIT{This is from QC07 paper.
We have verified that our structure is robustly stable
under MD, provided we maintain the $c=4.08$~\AA~ stacking
periodicity.  However, the outstanding effect of
relaxation in the real decagonals is ``puckering''~\cite{Gu-pucker}
whereby Al atoms bridging between Co atoms, as well
as those interior to HBS tiles,
rearrange their occupancy and displace from the planes,
modulated with a doubled ($2c$) periodicity.  
This is likely to {\it strengthen} the ``V-rule'', 
but might well flip the sign of the ``Fat/Fat'' matching rule.}


\WISHLIST{We discuss why the local configurations around Co found in these tiles
are optimal for the nearest-neighbor Al-Al and Al-Co interactions.}

\WISHLIST{The alternative route is in terms of the acceptance domains in
``perpendicular'' space, which (if sharply defined) confirm a
quasiperiodic structure.  This route is based on the following special
property of the potential in this layered structure:
the atom separations favored by the interaction are those
that also correspond to a small difference in perp-space coordinate.}

The matching rule depends on having the exact ratio of H:B:S tiles.
That is a worry: by local flips, one can trade tiles $HS \to BB$ or
vice versa.  Let's define energies $E_H, E_B, E_S$ as the sum
of interactions within each tile (including a chemical potential
$\mu_{\rm Al}$, since a flip changes the number of Al atoms). Then 
one naively expects to maximize or minimize $n_B$, depending whether
$E_H+E_S-2E_B$ is positive or negative, which would prevent
reaching the right composition. But the tile-tile (matching-rule)
interactions create a ``gap'' in Al site energies so that running the
reaction in {\it either} direction costs a positive energy.

The rule also depends on having the right internal Al decoration in 
every HBS tile.  We studied the same model with Al-Ni:
those potentials are quite similar to Al-Co (except
the Al-Ni attraction is not quite as strong as Al-Co).
The Penrose tiling may well be the ground state for Al-Ni
--- indeed, the {\it constrained} simulation behaves the same
for Al-Co and Al-Ni --- but it was less robust in the
{\it unconstrained} simulation: e.g.,
\OMIT{The Al corners with no Ni neighbor are less likely to be vacant.}
B tiles with just one internal Al on the mirror axis are metastable.
\OMIT{ this disrupts the matching rule between these Boats and adjacent tiles.
Lowering $\mu_{\rm Al}$ seems to convert $HS\to BB$ before it 
gets two Al on every B, however the half-constrained simulation found they
are not stable.}

As a reality check, we note that according to pair potentials, 
the (relaxed) matching-rule structure is unstable by 69.9 meV/atom 
compared to the tie-line 
between Al$_9$Co$_2$ (19\% Co) and other competing phases.
\OMIT{(some variant of) Al$_{13}$Co$_4$ (23.5\%Co)}
[For this calculation we used the optimal ``half-constrained''
simulation result for the same size as Fig.~\ref{fig:bestAlCo}.]
The (relaxed) ab-initio energy, using the VASP package~\cite{VASP},
came out unstable by 98.5 meV/atom.
\footnote{
After ab-initio relaxation, the cell expands 
by 1.20\% and 1.85\% along the long and short axes in-plane,
and by 1.96\% interlayer.
\OMIT{giving a net density 0.0662 atom/\AA$^3$, compared to
0.666 for Al$9$Co$_2$).}
The relatively small volume change supports the validity of
the pair potentials~\cite{Moriarty97}, which were conditioned  on a
particular electron density.}

\WISHLIST{An obvious change under relaxation
(by potentials or ab-initio) is the ring of 10 Al around 
a vacant corner, a decagon in projection converts into a star shape.}

For comparison, the best
Al-Co structure models~\cite{Mih-Wid-AlCo} are
$\approx 10$ meV/atom above the tie-line.  
In typical decagonal approximants, the energy is reduced
$\sim 50$meV/atom 
after Al atoms are allowed to ``pucker'' out of the layers
(doubling the $c$ axis to 8\AA), and to organize the proper
correlations between the puckerings of nearby atoms.
We have not yet investigated puckering in the present system, which 
would require molecular-dynamics simulation followed by 
relaxation~\cite{Gu-PML,Gu-pucker}.

We also attempted to make our structure more realistic by going to
a ternary, having recognized  that
the internal sites in HBS tiles (besides the interior vertex)
are always ``problem sites'' in Al-transition metal decagonals.
(E.g., those are the atoms that ``pucker''~\cite{Gu-pucker}.)
The root problem is that the two Al in one Boat are a bit overpacked,
in view of the Al-Al hardcore distance.  Could we fill these sites with
(somewhat smaller) Cu atoms?  No: when we tried an Al-Cu-Co ternary
(with pair potentials),
Cu atoms entered Hexagons forming CuCo pairs (as seen earlier in
Ref.~\cite{widom-match}.)

\vskip0pt plus2mm
\section[*] {Acknowledgements}
This work was supported by U.S. DOE grant
DE-FG02-89ER-45405; M.M. was also supported by Slovak
research grants VEGA 2/0157/08 and APVV-0413-06.
We thank A. Bhagat for discussions.
C.L.H. is grateful to the Slovak Academy of Sciences
for hospitality.

\end{document}